\documentclass[preprint,aps,superscriptaddress]{revtex4}
\usepackage{psfig}
\usepackage{amsmath,amssymb}
\begin{document}
\title{On the upper bound of the electronic kinetic energy in terms of density functionals.}
\author{L. Delle Site}
\affiliation{Max-Planck-Institute for Polymer Research, P.O. Box 3148,
  D-55021 Mainz, Germany.}
\begin{abstract}
We propose a simple density functional expression for the 
 upper bound of the kinetic energy for electronic systems.
Such a functional is valid in the limit of slowly varying density, 
its validity 
outside this regime is discussed by making a comparison with upper bounds obtained in previous work. The advantages of the functional proposed for applications to realistic systems is briefly discussed.
\end{abstract}
\maketitle
\section{Introduction}
Kinetic energy functionals of the electron density have been since long a subject of intense investigation. Starting with the pioneering work of March \cite{marchpr} through the monumental work of Lieb (see e.g. \cite{liebthirr,liebrmp1,lieblnp,liebrmp2}) the aim was to construct functionals which are accurate enough to properly describe physical and chemical properties and at the same time simple enough to allow a feasible computation. There has been a period of intense activity around this subject across the seventies until the end of the eighties during which many interesting results where produced. After about a  decade of weaker activity, the interest in the subject has got a new vigor (see e.g. the interesting work of Lude\~{n}a and coworkers \cite{ludenaa,ludena1,ludena2,ludena3} and the topical review of March \cite{marchdft}). The reason of this renewed interest lies in the fact that in the meanwhile novel computational schemes for quantum calculations, where the kinetic functional plays a key role, have been developed. Of particular interest is the linear scaling real space kinetic energy functional method, where the kinetic energy is calculated as a functional of the electron density. The electron wavefunctions are no more required and for this reason the method is called orbital-free density-functional theory (OFDFT) (see e.g.\cite{wang,watson,kaxiras,pino,tfc,kaxrev}). Since neither the diagonalization of the electronic Hamiltonian nor the reciprocal space sampling are required, such techniques allows for studies of relatively large systems compared to those treatable with the standard Kohn-Sham based approaches. Moreover, the fact that the calculation are done only in real space allows for the development of efficient quantum-classical interfaces which in current research are highly desirable within the emerging multiscale modeling techniques.
For this reason we turned the attention to the derivation of a simple and physically well founded kinetic functional. 
We start from the most general polar form of the electron wavefunction and derive an upper bound which is exact in the limit of slowly varying density; we discuss its validity beyond such an approximation by comparing our results with those available in literature.
A first interesting result is that our functional is potentially a better bound compared to some of those available in literature and by now well established; it is also universal, i.e. does not show explicit dependence on $N$, the number of electrons, differently from most of those found in literature. 
Next we combine our upper bound with the well known lower bound of Lieb and Thirring. By doing so, we conclude that a valid functional is likely to have the form of a Thomas-Fermi-Weizsacker type (with different constants), where the multiplicative constant of the Weizsacker term is the only free parameter.
To our knowledge this is the first time that a universal functional containing a Thomas-Fermi-like term and the Weizsacker one is obtained within the same derivation, and not by separated schemes. Moreover, the link we find between the wavefunction phase factor and the kinetic functional suggests an operative way to estimate the unknown constant.

\section{The Kinetic Energy}
Let us consider a system of $N$ electrons in a $3N$-dimensional volume $\Omega^{N}$. The general $N$-particle wavefunction in polar form is: $\psi({\bf r}_{1},{\bf r}_{2},....{\bf
  r}_{N})=\theta({\bf r}_{1},{\bf r}_{2},....{\bf r}_{N})
e^{i S({\bf r}_{1},{\bf r}_{2},....{\bf r}_{N})}$, where
$\theta({\bf r}_{1},{\bf r}_{2},....{\bf r}_{N})$ and $S({\bf
    r}_{1},{\bf r}_{2},....{\bf r}_{N})$ are real
functions in $\Omega^{N}$. We also require $\theta({\bf r}_{1},{\bf
  r}_{2},....{\bf r}_{N})$ to be antisymmetric with respect to any pair
permutation, i.e.
$\theta({\bf r}_{1},{\bf r}_{2},..{\bf r}_{i},..{\bf r}_{j}..{\bf r}_{N})=-
\theta({\bf r}_{1},{\bf r}_{2},..{\bf r}_{j},..{\bf r}_{i}..{\bf
  r}_{N})$
and $S({\bf r}_{1},{\bf r}_{2},....{\bf r}_{N})$ symmetric.
In this way the wavefunction $\psi({\bf r}_{1},{\bf r}_{2},....{\bf
  r}_{N})$ is antisymmetric as should be for a system of fermions.
For simplicity the spin variables are not explicitly considered.
We define the one electron density $\rho({\bf r})$ in $\Omega$ as:
$\rho({\bf r})=N\int_{\Omega^{N-1}}
\left[\theta({\bf r}_{1},...{\bf r}_{i-1},{\bf r}_{i},{\bf r}_{i+1},..{\bf
  r}_{N})\right]^{2}d{\bf r}_{1}..d{\bf r}_{i-1}d{\bf r}_{i+1}..d{\bf r}_{N}$,
where the index $i$ is arbitrary and can take any value from $1$ to
$N$; for the case $i=1$, $d{\bf r}_{i-1}$
is not considered while for $i=N$,
$d{\bf r}_{i+1}$ is not considered.
The one electron density, in turn, satisfies the condition:
$\int_{\Omega}\rho({\bf r})d{\bf r}=N$.
We will use atomic units $\hbar=1$, the electron mass $m=1$ and the electron charge $e=1$. Let us consider the average kinetic energy for the state $\psi$:
\begin{equation}
T_{\psi}=-\frac{1}{2}\int_{\Omega^{N}}\psi^{*}\nabla^{2}\psi d^{N}\omega
\label{eq1}
\end{equation}
where $d^{N}\omega=\Pi_{i=1}^{N}d{\bf r}_{i}$ and 
$\nabla=\sum_{i=1}^{N}\nabla_{i}$.
The expression above can be written also as (see e.g. \cite{liebrmp1}):
\begin{equation}
T_{\psi}=\frac{1}{2}\int_{\Omega^{N}}|\nabla\psi|^{2}d^{N}\omega.
\label{eq2}
\end{equation}  
Substituting the expression $\psi=\theta e^{iS}$ into Eq.\ref{eq2} one obtains:
\begin{equation}
T_{\psi}=\int_{\Omega^{N}}\left[\frac{\theta^{2}|\nabla
  S|^{2}}{2}+\frac{|\nabla\theta|^{2}}{2}\right] d^{N}\omega.
\label{eq3}
\end{equation}
The integrand on the r.h.s. of Eq.\ref{eq3} is a $3N$-dimensional function; the goal is to reduce it, as rigorously as possible, to a simple three dimensional functional of $\rho({\bf r})$. 
\subsection{Upper bound to $|\nabla S|$ in the slowly varying density limit}
The average over $\Omega^{n}$ of the $3N$-dimensional momentum vector of the system is defined as:
\begin{equation}
\left<\psi|{\bf m}|\psi|\right>_{\Omega^{N}}=-  \int_{\Omega^{N}}
Im\left[\psi^{*}\nabla\psi\right]d^{N}\omega
\label{eq4}
\end{equation}
substituting the polar form of $\psi$ into Eq.\ref{eq4} we obtain:
\begin{equation}
  \int_{\Omega^{N}}
Im\left[\psi^{*}\nabla\psi\right]d^{N}\omega=
  \int_{\Omega^{N}}\left[\psi^{*}\nabla S\psi\right]d^{N}\omega.
\label{eq5}
\end{equation}
Through Eq.\ref{eq5} we find the relation $\left<\psi|{\bf m}|\psi\right>_{\Omega^{N}}=\left<\psi|\nabla
S|\psi\right>_{\Omega^{N}}$, which suggests an interpretation of  $\nabla S$ as a $3N$-dimensional quantum velocity field. Such an interpretation would not be new, and can be often found in literature (see e.g. \cite{salesi,hu,shimbori}), above all in the context of a fluid dynamics formulation of quantum mechanics \cite{gd}. If we interpret $\nabla S$ as a quantum velocity field than, in the limit of $\rho({\bf r})$ being a slowly varying function, we can state the following inequality for the absolute value of $\nabla S$:
\begin{equation}
\left<\psi|{\bf |}\nabla S{\bf |}|\psi\right>\leq \left< \psi||{\bf P}_{F_{N}}||\psi\right>
\label{eq6}
\end{equation}
which explicitly reads:
\begin{equation}
\int_{\Omega^{N}}\theta^{2}|\sum_{i=1}^{N}\nabla S| d^{N}\omega\leq
\int_{\Omega^{N}}\theta^{2}|\sum_{i=1}^{N}{\bf P}_{F_{i}}|d^{N}\omega.
\label{eq8}
\end{equation}
Where ${\bf P}_{F_{i}}={\bf P}_{F}({\bf r}_{i})$, with ${\bf
    P}_{F_{N}}=\sum_{i=1}^{N}{\bf P}_{F}({\bf r}_{i})$,  
is the Fermi momentum, i.e. the maximum momentum a
particle can reach in the limit of slowly varying density.
Since Eq.\ref{eq8} must hold for any arbitrary 
subvolume of $\Omega^{N}$ we obtain:
\begin{equation}
\left|\nabla S\right|\leq \left|{\bf P}_{F_{N}}\right|; \forall {\bf R}\in \Omega^{N}
\end{equation}
and thus
\begin{equation}
\left|\nabla S\right|^{2}\leq\left|{\bf P}_{F_{N}}\right|^{2}; \forall {\bf
  R}\in \Omega^{N}.
\label{eq9}
\end{equation}
From the inequality \ref{eq9} we have:
\begin{equation}
\int_{\Omega^{N}}\left(\psi^{*}\left|\nabla S\right|^{2}\psi\right) d^{N}\omega\leq
\int_{\Omega^{N}}\left(\psi^{*}\left|{\bf P}_{F_{N}}\right|^{2}\psi\right)d^{N}\omega
\label{eq10}
\end{equation}
or equivalently:
\begin{equation}
\int_{\Omega^{N}}\theta^{2}|\nabla S|^{2} d^{N}\omega\leq
\int_{\Omega}\rho({\bf r})\left|{\bf P_{F}}({\bf r})\right|^{2}d{\bf r}.
\label{eq11}
\end{equation}
The 
second term on the r.h.s. of Eq.\ref{eq11} is obtained in the
following way:
\begin{equation}
\int_{\Omega^{N}}\left(\psi^{*}\left|{\bf P}_{F}\right|^{2}\psi \right)d^{N}\omega=
\sum_{i=1}^{N}\int_{\Omega_{i}}\left|{\bf P_{F_{i}}}\right|^{2}d{\bf r}_{i}\int_{\Omega^{N-1}}\theta^{2}d^{N-1}\omega=\sum_{i=1}^{N}\int_{\Omega_{i}} \frac{\rho({\bf r}_{i})}{N}\left|{\bf P_{F_{i}}}\right|^{2}d{\bf r}_{i}.
\label{eq12}
\end{equation}
If we go back to Eq.\ref{eq3}, since:
$|{\bf P}_{F}({\bf r})|=C_{F}[\rho({\bf r})]^{1/3}$, 
where $C_F$ is a constant, $(C_F)^{2}=(3\pi^{2})^{2/3}$, 
we have:
\begin{equation}
\int_{\Omega^{N}}\theta^{2}\frac{|\nabla
    S|^{2}}{2}d^{N}\omega +\int_{\Omega^{N}}\frac{1}{2}\sum_{i=1}^{N}|\nabla_{i}\theta|^{2}d^{N}\omega\leq\int_{\Omega}\frac{C_{F}^{2}}{2}[\rho({\bf r})]^{5/3}+\int_{\Omega^{N}}\frac{1}{2}\sum_{i=1}^{N}|\nabla_{i}\theta|^{2}d^{N}\omega
\label{eq12b} 
\end{equation}
In this way we have determined an upper bound to $T_{\psi}$, where the first
term is written as a functional of $\rho({\bf r})$. 
We shall now reduce also the other term into a three dimensional expression,
possibly as a functional of $\rho({\bf r})$. One may wonder why to study the phase factor $S$ for atoms and molecules in their ground state. Indirectly this work suggests an unconventional answer; strictly speaking $S=0$ for the ground state of atoms and molecules,
and this would mean that, in principle, the only kinetic functional term is the Weizsacker term, as it is discussed by Sears {\it et al.} \cite{sears}, Herring \cite{herr} and Luo \cite{luo}. However in Refs.\cite{herr,luo} is discussed and shown that this results would not be correct and there must be an additional term to the kinetic energy coming from some angular part of the wavefunction which factorizes, i.e. something related to a phase factor. At the same time if $\psi=\theta e^{iS}$ is a solution, also $\psi^{*}=\theta e^{-iS}$ is a solution with the same energy; this also holds for linear combinations of the two.
For atoms and molecules one can always choose, without loss of generality, linear combinations of $\psi$ and $\psi^{*}$ that lead to real wavefunctions. In this case, as it can be verified by a straightforward calculation, the r.h.s. of Eq.\ref{eq12b} is still an upper bound to the kinetic energy.
\subsection{Weizsacker and non local Information functional}
The term 
$\int_{\Omega^{N}}|\nabla\theta|^{2}d^{N}{\bf r}=\sum_{i=1}^{N}\int_{\Omega^{N}}|\nabla_{i}\theta|^{2}d^{N}{\bf r}$
is what Sears {\it et al.} ~\cite{sears} refer to as the {\it multivariate kinetic
functional} and can be written as:
\begin{equation}
\frac{1}{2}\sum_{i=1}^{N}\int_{\Omega^{N}}|\nabla_{i}\theta|^{2}d^{N}{\bf r}=
\frac{1}{8}\int_{\Omega}\frac{|\nabla\rho({\bf r})|^{2}}{\rho({\bf
    r})}d{\bf r}+\frac{1}{8}\sum_{i=1}^{N}\int_{\Omega^{N}}\frac{\rho({\bf
  r}_{i})}{N}\frac{|\nabla_{i}f({\bf r}_{1},...{\bf r}_{i-1},{\bf
    r}_{i+1}...{\bf r}_{N}/{\bf r}_{i})|^{2}}{f({\bf r}_{1},
...{\bf r}_{i-1},{\bf r}_{i+1}...{\bf r}_{N}/{\bf r}_{i})}d^{N}{\bf r}
\label{eq13}
\end{equation}
where the first term on the r.h.s. is the well known Weizsacker term and $i\in \left[1,N\right]$ with $f({\bf r}_{1},
...{\bf r}_{i-1},{\bf r}_{i+1}...{\bf r}_{N}/{\bf r}_{i})$ is
proportional to the conditional density, i.e. the probability density 
of
finding a certain spatial configuration for the $N-1$ particles
once the position of the $i$-th particle is assigned (see also \cite{kohout}).
Following the work of Ref.\cite{sears} we can write:
\begin{equation}
\frac{1}{2}\sum_{i=1}^{N}\int_{\Omega^{N}}|\nabla_{i}\theta|^{2}d^{N}{\bf r}=\frac{1}{8}\int_{\Omega}\frac{|\nabla\rho({\bf r})|^{2}}{\rho({\bf r})}d{\bf r}+\frac{1}{8}\int_{\Omega}\rho({\bf r})I({\bf r})d{\bf r}
\label{eq14}
\end{equation}
where
\begin{equation}
\int_{\Omega}\rho({\bf r})I({\bf r})d{\bf r}=\int_{\Omega}\rho({\bf
  r})\left[\int_{\Omega^{N-1}}\frac{|\nabla_{i}f({\bf r}_{2},{\bf
      r}_{3...}.{\bf r}_{N}/
{\bf r})|^{2}}{f({\bf r}_{2},{\bf r}_{3},...{\bf r}_{N}/{\bf r})}d{\bf 
r}_{2}d{\bf r}_{3}...d{\bf r}_{N}\right]d{\bf
r}
\label{eq15}
\end{equation}
with $I({\bf r})$ being the well known non local {\it information functional} within Fisher information theory. An exact expression for $I({\bf r})$ is difficult to find, however one can notice that $I({\bf r})$ would be equivalent to write in local form some kinetic correlation effects which are usually considered negligible \cite{ludena1}.
Combining the results above with those of the previous section we obtain:
\begin{equation}
T_{\psi}\le \frac{[C_{F}]^{2}}{2}\int_{\Omega}[\rho({\bf r})]^{5/3}+\frac{1}{8}\int_{\Omega}\frac{|\nabla\rho({\bf r})|^{2}}{\rho({\bf
r})}d{\bf r}+\frac{1}{8}\int_{\Omega}\rho({\bf r})I({\bf r})d{\bf r}
\label{eq16}
\end{equation} 

\section{Beyond the slowly varying density regime}
In this section, by comparing the result of Eq.\ref{eq16} with upper bounds to $T_{\psi}$ available in literature, we will discuss the validity of our results for the general case. For simplicity we will neglect the non local information functional term since, as said before, it represents a non relevant correction. Under this hypothesis Eq.\ref{eq16},
becomes:
\begin{equation}
T_{\psi}\le \frac{[C_{F}]^{2}}{2}\int_{\Omega}[\rho({\bf r})]^{5/3}+\frac{1}{8}\int_{\Omega}\frac{|\nabla\rho({\bf r})|^{2}}{\rho({\bf
r})}d{\bf r}.
\label{eq18}
\end{equation}
It is encouraging to note that the same functional form for the upper bound, with a slightly different multiplicative constant for the first term, was conjectured, following arguments different from ours, by Lieb \cite{lieblnp}. March and Young obtained a result similar to ours but the constant multiplying the $\int \rho^{5/3} d{\bf r}$ must be determined for each atom \cite{marchpr}. Gazquez and Robles \cite{garo} derived a kinetic energy functional of the form: $T_{\psi}=C_{1}\left(1-\frac{C_{0}}{N^{1/3}}\right)\int \rho^{5/3} d{\bf r}+\frac{1}{8}\int \frac{|\nabla\rho|^{2}}{\rho}d{\bf r}$, the same functional was independently proposed by Acharya {\it et al.} \cite{absp}.
This functional has explicit dependence on $N$, the number of particle, thus it is not universal, however, in the thermodynamic limit, i.e. $N$ large, has the same functional form of the functional we derived (with, again, a slightly different constant for the first term). Moreover our result is fully consistent with the estimate of the {\it relative-phase-energy} term of Herring \cite{herr} where he concludes that the upper bound must be something which has the same form as our functional. 
Even better comparison can be made for, the simpler, one dimensional case; now the term $\int \rho^{5/3} d{\bf r}$ in Eq.\ref{eq18} becomes $\int \rho^3 d{\bf x}$ because in our derivation $P_{F}\sim \rho$ in one dimension. The resulting upper limit 
has the same functional form for $T_{\psi}$ obtained by Harriman \cite{harr} using the special equidensity orbitals (SEDOs) construction for $\psi$ (see also \cite{macke}).
Moreover it represents an upper bound, in the limit of large $N$, 
to the rigorous functional found by March and Young in one dimension: $T_{\psi}\le const\times\int \rho^{3/2}dx+\int \frac{|\nabla\rho|^{2}}{\rho}dx$. 
The arguments above, although do not represent an explicit proof, suggest that the functional we propose may indeed be a valid upper bound beyond the slowly varying density approximation.
An intuitive argument to support the validity of our hypothesis is the following. In the  slowly varying density regime, our upper bound condition on the momentum of the single electron is rather ''large''. In fact we approximate the momentum of each electron with the maximum value possible instead of distributing  among the $N$ electrons all the states available from $|{\bf m}|=0$ to $|{\bf m}|=|{\bf P}_{F}|$. By slowly moving from this regime to an intermediate one, for a large number of particles, despite the fact that ${\bf P}_{F}$ slowly looses its physical meaning, the upper bound hypothesis $|\nabla S|\le const\times\rho^{1/3}$ it is likely to still hold for an extended range of densities.
Moreover, the larger $N$ the more extended the range of validity; in the thermodynamic limit one can expect such an upper limit to be always valid, as the comparison with the functional of Gazquez and Robles \cite{garo} and of  Acharya {\it et al.} \cite{absp} suggests.
Interestingly, if this was the case, we would have found a better upper bound to $T_{\psi}$ compared to that of Zumbach\cite{zum1}: 
\begin{equation}
T_{\psi}\le [1+C_{Zu}N^{2/3}]\frac{1}{8}\int \frac{|\nabla\rho|^{2}}{\rho}d{\bf r};~~~~C_{Zu}=15(4\pi)^{2}\frac{3}{5}\left(\frac{1}{5}\right)^{2/3}
\label{eq19}
\end{equation}
which was obtained using a  SEDOs-like construction of $\psi$ and is considered of general validity, although its explicit dependence on $N$ makes it not universal. In fact, as it is discussed by Pathak and Gadre \cite{pathgadr}, and also reported by Zumbach \cite{zum1}, by using Schwartz inequality and then applying one Sobolev inequality in three dimensions it is possible to obtain the following relation:\begin{equation}
\int\rho^{5/3}d{\bf r}\le C_{PG}\times N^{2/3}\frac{1}{8}\int \frac{|\nabla\rho|^{2}}{\rho}d{\bf r}
\label{eq20}
\end{equation}
where $C_{PG}$ is a constant (see \cite{pathgadr}), if now we compare Eq.\ref{eq18} with Eq.\ref{eq19} via Eq.\ref{eq20}, we obtain:
\begin{equation}
T_{\psi}\le \frac{[C_{F}]^{2}}{2}\int_{\Omega}\rho^{5/3}+\frac{1}{8}\int_{\Omega}\frac{|\nabla\rho|^{2}}{\rho}d{\bf r}\le  [1+C_{Zu}N^{2/3}]\frac{1}{8}\int \frac{|\nabla\rho|^{2}}{\rho}d{\bf r}
\label{eq21}
\end{equation}
since $C_{PG}\times C_{F}^{2}\ll C_{Zu}$.
\section{Upper and Lower Bound for $T_{\psi}$}
An interesting property of our upper bound can be obtained by relating it to the Lieb-Thirring inequality and to the consequent lower bound to $T_{\psi}$ \cite{liebrmp1,liebthirr,liebrmp2,spruch,daub}:
\begin{equation}
T_{\psi}\ge \frac{C_{LT}}{2}\int\rho^{5/3}d{\bf r}.
\label{eq22}
\end{equation}   
From Refs.\cite{liebrmp1,spruch,daub} it follows that $C_{F}^{2}$ of Eq.\ref{eq18} is very close to $ C_{LT}$ of Eq.\ref{eq22}. In fact $C_{F}^{2}\approx 9.57$ while $C_{LT}\approx 9.11$. Actually Lieb in Ref.\cite{liebrmp1} argues that numerical calculations improve $C_{LT}$ to $9.578$. If we consider 
 $C_{F}^{2}\approx C_{LT}$, then we can write in good approximation the following relation:
\begin{equation}
C\times\int\rho^{5/3}d{\bf r}\le T_{\psi}\le C\times\int\rho^{5/3}d{\bf r}+\frac{1}{8}\int_{\Omega}\frac{|\nabla\rho|^{2}}{\rho}d{\bf r}
\label{eq23}
\end{equation}   
where $C$ is a constant whose value is between that of $C_{F}^{2}/2$ and $C_{LT}/2$, thus very close to both. The implications of Eq.\ref{eq23} are very interesting; it suggests that a valid approximation for $T_{\psi}$ can be written as:
\begin{equation}
T_{\psi}\approx C\times\int\rho^{5/3}d{\bf r}+q\times \frac{1}{8}\int_{\Omega}\frac{|\nabla\rho|^{2}}{\rho}d{\bf r}
\label{eq24}
\end{equation}
where $0\le q \le1$, is the only free parameter. 
A possible way to obtain a first guess for $q$ would be by estimating $\nabla S$ numerically for some simple model systems. However already at this stage Eq.\ref{eq24} represents a very interesting result, as it has been previously discussed in the introduction, in connection with current quantum based computational schemes.
\section{Discussion and Conclusions}
We have used the polar form of a many-particle electron wavefunction to derive a simple functional for the kinetic energy. Such a functional represents an upper bound to the true kinetic energy and it is exact in the limit of slowly varying density. We have discussed its validity beyond such an approximation by making a comparison with functionals obtained in previous work. The novelty of our derivation presents different aspects; we obtain within the same theoretical framework a  kinetic functional containing both a Thomas-Fermi like functional and the Weizsacker functional. These are usually determined by following separated procedures  under different physical approximations. This result provides a physical justification to a high desirable upper bound functional which was heuristically conjectured by Lieb \cite{lieblnp}. Another advantage is that it does not show explicit dependence on $N$ , differently from most of the upper bounds available in literature, and represents a better bound compared to other well established functionals. Finally, combined with the Lieb-Thirring lower bound leads to the conclusion that a valid functional, which can well approximate the true one, could be determined by simply tuning the multiplicative constant (between zero and one) of the  Weizsacker term. This result is 
very interesting above all for applications to condensed matter systems
within free orbital density functional schemes. In fact, it provides not only a theoretical background to justify the kinetic functional currently employed, but also a possible procedure to determine new ones. 
Approaches as the one shown in this paper, are crucial for the development of computationally efficient and theoretical flexible quantum mechanical techniques for modern multiscale simulations . As discussed into the introduction, the computational apparatus is available and new ideas, which can improve the energy functionals currently available or suggest new ones, are strongly required. In this respect, our contribution suggests a possible way to proceed.
\section{Acknowledgments}
I would like to thank professor Elliot H.Lieb  for his useful remarks.

\end{document}